# Field-free nucleation of antivortices and giant vortices in non-superconducting materials

Morten Amundsen,[1] Jabir Ali Ouassou,[1] and Jacob Linder[1]

[1]*Center for Quantum Spintronics, Department of Physics, Norwegian University of Science and Technology, NO-7491 Trondheim, Norway*

Giant vortices with higher phase-winding than $2\pi$ are usually energetically unfavorable, but geometric symmetry constraints on a superconductor in a magnetic field are known to stabilize such objects. Here, we show via microscopic calculations that giant vortices can appear in intrinsically non-superconducting materials, even without any applied magnetic field. The enabling mechanism is the proximity effect to a host superconductor where a current flows, and we also demonstrate that antivortices can appear in this setup. Our results open the possibility to study electrically controllable topological defects in unusual environments, which do not have to be exposed to magnetic fields or intrinsically superconducting, but instead display other types of order.

*Introduction.* It is well-known that applying a magnetic field to a type-II superconductor can lead to the formation of Abrikosov vortices [1]. A gradient in the phase $\varphi$ of the superconducting order parameter $\Delta = |\Delta|e^{i\varphi}$ causes a circulating supercurrent around such vortices, whereas $|\Delta| \to 0$ at their centers. Vortex excitations in superconductors [2, 3] remains a vibrant research topic, not least because it lies at the intersection of two major research fields: superconductivity and topology in physics.

It was recently pointed out in Ref. 4 that it is also possible to generate Josephson vortices without applying magnetic fields. Such vortices are also characterized by a quantized phase-winding and a suppressed order parameter at their core [5]. Motivated by this, we have performed microscopic calculations using the quasiclassical theory of superconductivity [6] on a normal metal enveloped by a current-biased superconducting wire (Fig. 1). The idea behind the device is simple: an external current source forces a supercurrent through the wire, and this circulation whirls the condensate in a proximitized normal metal as well. Our objective has been to determine what type of electrically controllable vortex physics that then emerges. *We demonstrate here that both giant vortices and antivortices appear in the non-superconducting region even in the absence of any applied magnetic field.* This provides an alternative method of creating complex vortex patterns by applying electric currents. Since these patterns are generated in a proximitized non-superconductor, this opens up the intriguing prospect of studying unusual topological vortex excitations in materials with other types of quantum order, which do not have to be compatible with bulk superconductivity. One example is a magnetic metal, where the generation of odd-frequency triplet superconducting order could reverse the chirality of some vortices, similarly to the paramagnetic Meissner effect [7, 8]. More fundamentally, it raises the intriguing question: what characterizes a vortex in an odd-frequency order parameter?

*Geometric effect and winding-number.* Since a circulating supercurrent requires a finite phase-gradient $\nabla\varphi$, and the analyticity of the superconducting wave function implies integral winding numbers $n = \Delta\varphi/2\pi$ around any point, the system is topologically coerced into nucleating vortices in the normal metal region of Fig. 1. Assuming a thin-film structure, the total charge current associated with this circulation is small, and the magnetic field generated by the circulation can safely be neglected. Note that in contrast to the setup proposed in Ref. 4, our normal metal is surrounded by a continuous superconducting wire on all sides, instead of having two separate wires on the top and bottom, which we will show fundamentally alters the vortex physics in the system. Another important difference is that we model the superconducting wire using an exact solution of the Usadel equation in the current-biased superconductor and tunneling boundary conditions. As we will demonstrate, this is necessary to correctly describe qualitative changes that the phase-winding induces in *e.g.* the density of states (DOS).

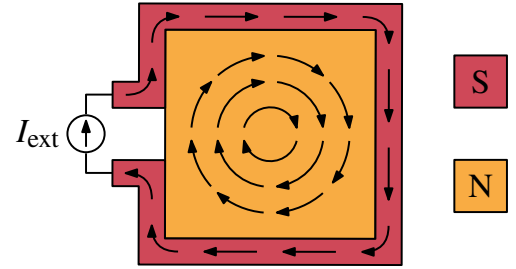

FIG. 1: Conceptual sketch of the physical system. An external current source is used to inject a current into a superconductor (red). The circulating current also affects a proximitized normal metal (yellow), causing an electrically controlled vortex to emerge there.

When the current in the superconducting wire makes a total winding number $N > 1$, there are multiple ways to satisfy the boundary conditions. Among other possibilities, we can get (*i*) $N$ vortices with a winding 1 each, (*ii*) $N + M$ vortices with a winding $+1$ and $M$ antivortices with winding $-1$, or (*iii*) just one giant vortex with a winding $N$. The kinetic energy of an $n$-winding vortex scales with $n^2$, so the most energetically favorable is configuration (*i*). Hence, giant vortices and antivortices are seldom seen. However, since the superconducting order parameter respects the symmetries of the underlying geometry, vortices only nucleate along the symmetry axes of the system. For highly symmetric geometries, these additional constraints may force the appearance of giant vortices or antivortices. The resulting interplay between topological defects, geometric symmetries, and energy minimization was previously studied in Refs. [9–12] using the phenomenological Ginzburg-Landau formalism for type-II superconductors in a magnetic field. Here, we show that this effect also arises in proximitized normal metals without magnetic fields. This generalization is particularly important as it opens the possibility to study novel vortex physics in materials featuring completely different order than superconductors, *e.g.* ferromagnets or topological insulators.

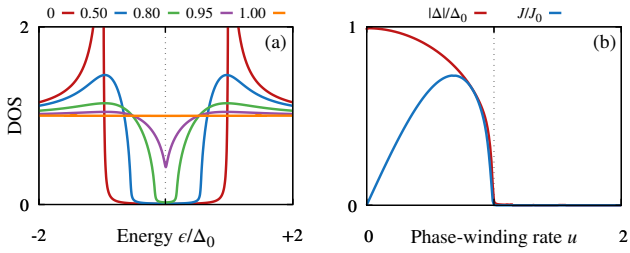

FIG. 2: Visualization of a bulk superconductor with a uniform current at zero temperature. (a) DOS for varying winding rates $u$, as shown in the legends above. Note how the coherence peaks are smoothed out for $u > 0$ and the gap closes as $u \to 1$, illustrating a qualitatively different behaviour for $u > 0$. (b) Gap $\Delta$ and current density $J$ as functions of $u$. As long as $u < 1/2$, we see that $\Delta \approx \Delta_0$ and $J \sim u$, and a non-selfconsistent solution is reasonably accurate. However, the current becomes non-monotonic for $u > 2/3$, so this regime is inaccessible in our proposed experimental setup.

*2D diffusive metal with phase-winding.* As shown in Fig. 1, we consider a normal metal with a superconducting loop grown on top. We describe the properties of the metal in terms of quasiclassical propagators $\hat{g}$ in Nambu and spin space,

$$\hat{g}(\bm{r},\epsilon) = \begin{pmatrix} g(\bm{r},+\epsilon)\sigma_0 & f(\bm{r},+\epsilon)i\sigma_2 \\ -f^*(\bm{r},-\epsilon)i\sigma_2 & -g^*(\bm{r},-\epsilon)\sigma_0 \end{pmatrix}, \quad (1)$$

where the normal part $g$ and anomalous part $f$ are complex scalar functions, subject to the normalization constraint $\hat{g}^2 = 1$. Here, $\sigma_0$ is the $2 \times 2$ identity matrix, and $\sigma_2$ is the second Pauli matrix. We assume that all length scales in the problem are large compared to the Fermi wavelength and mean free path, *i.e.* we take the quasiclassical diffusive limit. The propagators $\hat{g}$ are then governed by the Usadel equation [6, 13, 14],

$$D\nabla(\hat{g}\nabla\hat{g}) + i[\epsilon\hat{\tau}_3 \,,\, \hat{g}] = 0, \quad (2)$$

where $D$ is the diffusion constant, $\epsilon$ the quasiparticle energy, and $\hat{\tau}_3 = \mathrm{diag}(+\sigma_0, -\sigma_0)$. Furthermore, we assume that the normal region is connected to the superconducting wire by a low-transparency interface. We may then use the Kupriyanov-Lukichev boundary condition $\zeta \bm{e}_\perp \cdot \nabla \hat{g}_n = [\hat{g}_n, \hat{g}_s]/\xi$ [15], where $\zeta$ parametrizes interface resistance, $\bm{e}_\perp$ is the outwards-pointing interface normal vector, $\hat{g}_n$ and $\hat{g}_s$ are propagators on the normal and superconducting sides, and $\xi$ the superconducting coherence length. The propagators $\hat{g}_s$ in the current-biased superconductors were evaluated analytically. The applied current also creates a magnetic field which penetrates the proximitized material. Its strength depends on the total applied current, which in turn depends on the pair density and dimensions of the superconductor. However, since the field is perpendicular to and roughly constant within the current loop, its only effect is to slightly perturb the applied current for which a given vortex pattern appears. We have neglected the quantitative correction from the magnetic field herein.

In practice, the differential equations above are Riccati-parametrized for stability [16], and then solved numerically using a finite-element method on a two-dimensional mesh. This lets us handle arbitrary sample geometries, such as the regular polygons considered herein. For more information about the numerical solution procedure itself, see Ref. 17.

*Superconducting wire with a uniform current.* As shown in the supplemental (Sec. II), the propagator $\hat{g}$ in a current-biased bulk superconductor can be written [18, 19]

$$\hat{g} = \frac{1}{\sqrt{\epsilon^2 - \Theta^2}} \begin{pmatrix} +\epsilon\,\sigma_0 & \Theta\,e^{+i\varphi}\,i\sigma_2 \\ \Theta\,e^{-i\varphi}\,i\sigma_2 & -\epsilon\,\sigma_0 \end{pmatrix}, \quad (3)$$

where $\Theta(\epsilon)$ parametrizes the strength of the superconductivity, and $\varphi$ is the superconducting phase. The phase varies linearly with the distance $\ell$ along the wire. Defining $\varphi(0) \equiv 0$, and parametrizing the variation using a winding rate $u \equiv \xi|\nabla\varphi|$, we therefore get $\varphi(\ell) = u\ell/\xi$. The function $\Theta(\epsilon)$ is determined by:

$$\Theta = \frac{|\Delta|}{1 + u^2/2\sqrt{\Theta^2 - \epsilon^2}}, \quad (4)$$

$$|\Delta| = \frac{1}{\mathrm{acosh}\,\omega_c} \int_0^{\omega_c} d\epsilon\,\mathrm{Re}\left(\frac{\Theta}{\sqrt{\epsilon^2 - \Theta^2}}\right) \tanh\left(\frac{\pi}{2e^\gamma}\frac{\epsilon}{T}\right). \quad (5)$$

These equations have been written in a form where $\Theta, \Delta, \epsilon, \omega_c$ are all normalized to the zero-current gap $\Delta_0$, while the temperature $T$ is normalized to the critical temperature $T_c$. Here, $\omega_c$ refers to the Debye cutoff, and $\gamma$ is the Euler-Mascheroni constant. The first of these equations is a fixpoint iteration equation. This is easily solved by guessing $\Theta(\epsilon) = 1$ and $|\Delta| = 1$, and applying Newton's method to the equation for a discretized set of energies from the Debye cutoff $\epsilon = \omega_c$ to zero energy $\epsilon = 0$. The second is a selfconsistency equation for the gap $\Delta$, which is evaluated by numerical integration of the results for $\Theta(\epsilon)$. We then alternate between solving the fixpoint equation and selfconsistency equation until satisfactory convergence. The solutions to the equations above are visualized in Fig. 2.

When approaching the setup in Fig. 1 numerically, we assumed that the superconducting wire suffers a negligible inverse proximity effect from the normal metal. In this case, we can use the analytical equation above for the superconducting wire, and reduce the superconductor to effective boundary conditions for the normal metal. Furthermore, we numerically only considered phase-winding rates $u \leq 0.5$, in which case Eq. (5) can be replaced by the approximation $|\Delta| \approx 1$. Note that since the phase-winding rate $u$ cannot be arbitrarily large, we need a system much larger than the coherence length to obtain high winding numbers using a current bias.

*Quantifying vortices.* We can study the proximity-induced superconductivity in a normal metal via the pair-correlation

$$\Psi(\bm{r}) \sim \int_0^{\omega_c} d\epsilon\,[f(\bm{r},+\epsilon) - f(\bm{r},-\epsilon)]\tanh(\epsilon/2T), \quad (6)$$

which behaves like a complex order parameter. This pair-correlation can be decomposed as $\Psi \equiv |\Psi|\exp(i\varphi)$, and the phase $\varphi$ can then be extracted using $\varphi = \arctan(\mathrm{Im}\,\Psi/\mathrm{Re}\,\Psi)$.

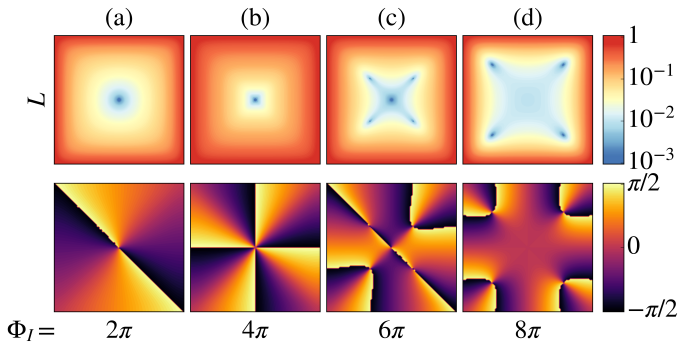

FIG. 3: Vortex nucleation patterns for various applied current windings $\Phi_I$, for a quadratic normal metal with side lengths $L = 12\xi$. The top row shows the magnitude $|\Psi|$ of the pair correlation, where the minima indicate the vortices. The bottom row shows the phase $\varphi$ of the pair correlation, from which the winding of individual vortices can be determined. The total windings $\Phi_I$ are listed below.

As discussed in the introduction, the circulating current in the enclosing superconductor creates a phase-winding $\nabla\varphi$ along the interface. However, the phase $\varphi$ is uniquely defined modulo $2\pi$, which means that it is only possible for the phase to vary continuously around the edges of the normal metal if it increases by $\Phi_I = 2\pi N$ after having traversed the entire circumference. In other words, we must have a total vorticity

$$N = \frac{\Phi_I}{2\pi} \equiv \frac{1}{2\pi}\oint_{\partial\Omega}(\nabla\varphi)\cdot d\boldsymbol{\ell}, \quad (7)$$

where $\partial\Omega$ is the boundary of the normal metal. When we have a finite vorticity $N$, the currents inside the normal metal will form closed loops, leading to the appearance of vortices. More precisely, the total vorticity $N$ will be equal to the sum of the winding numbers $n$ of all the induced vortices. The vortices manifest as nodes in the pair correlation $\Psi$.

*Numerical results.* In the upper row of Fig. 3, the vortex pattern for increasing applied current winding $\Phi_I$ is shown. The winding of the individual vortices may be determined graphically from the phase of the pair correlation function $\varphi$, which is plotted in the bottom row of Fig. 3. By using Eq. (7) with the replacements $N \to n$ and $\partial\Omega \to \mathcal{C}$, where $\mathcal{C}$ is any contour encircling a single vortex, one sees that $n \neq 0$ only if the integration path crosses discontinuities. Furthermore, each discontinuity contributes a value to the integral equal to the size of the jump. For $\Phi_I = 2\pi$, shown in Fig. 3(a), there is a single vortex in the center of the normal metal, and any closed contour around this point must traverse two jumps $\Delta\varphi = \pi$, thus showing that the vortex has a winding $n = +1$. We note that the precise locations of these discontinuities depend on the reference point for the phase of the superconductors, and are hence not physically significant. The number of times a closed loop crosses a discontinuity, however, is. In Fig. 3(b), where $\Phi_I = 4\pi$, there is still only a single vortex in the system, but now the plot of $\varphi$ shows four discontinuities, from which it is inferred that this is a giant vortex with $n = +2$.

For $\Phi_I = 6\pi$, shown in Fig. 3(c), five vortices are found. As the sum of the individual topological numbers should add up to $N = +3$, in accordance with Eq. (7), one of these vortices must be an antivortex. The phase plot shows that this is indeed the case: the central vortex winds in the opposite direction of other vortices. Hence, this configuration consists of one central $n = -1$ antivortex with four surrounding $n = +1$ vortices. For $\Phi_I = 8\pi$, there are four regular $n = +1$ vortices along the diagonals, as shown in Fig. 3(d). Since these vortex patterns arise from symmetry constraints, they are naturally sensitive to asymmetries in the geometry. The giant vortex in Fig. 3(b) splits into two $n = +1$ vortices as the geometry becomes rectangular. However, the vortices continue to overlap strongly for sufficiently small deviations, as shown in Sec. III of the supplemental. This means that the giant vortex could in practice be stabilized against deviations from perfect symmetry by creating a pinning potential at this location [20]. Since the vortex positions are also influenced by the applied currents, another option is to fine-tune the currents to experimentally realize the giant vortex. The pattern in Fig. 3(c) is, on the other hand, stable against small deviations in aspect ratio. The reason is that when the geometry becomes increasingly rectangular, it eventually becomes energetically favorable to satisfy $N = +3$ as three $n = +1$ vortices along the longest axis. The transition to such a pattern can only occur in a way which respects the symmetries of the rectangle, and hence the central antivortex turns into a vortex, and additional antivortices must appear so that the off-center vortices can annihilate symmetrically [21].

The vortices also create a spatial modulation of the DOS: at the vortex cores, superconductivity vanishes, and the minigap disappears. This means that the vortices we predict can be directly inferred via local STM measurements. In Fig. 4, the DOS for $\epsilon = 0$ is plotted along the diagonal of the normal metal (*i.e.*, between two opposite corners). This confirms that the normal-state result DOS = 1 is recovered at the vortex cores. For the $n = +2$ vortex produced by $\Phi_I = 4\pi$, the minigap is suppressed in a larger region around the vortex than for $\Phi_I = 2\pi$. For $\Phi_I = 6\pi$, the normal region is larger still, but this is likely due to the close proximity of three vortices. For $\Phi_I = 8\pi$, the vortices are sufficiently far apart for a dip in the DOS to appear inbetween, providing an observable signature.

The above can be understood by analyzing the pair correlation. In the supplemental (Sec. I), it is shown that for small distances $r$ from the vortex center, $\Psi \sim (r/2\xi_0)^n/n!$, where $\xi_0$ is the Ginzburg-Landau coherence length. For $r < 2\xi_0(n!)^{1/n} \approx 2[1 + (n-1)/e]\xi_0$, these correlations recover more slowly with increasing winding $n$, and hence the minigap is increasingly suppressed. The fact that the vortex size increases linearly with $n$ also in the diffusive limit can be motivated from Fig. 2. There, we see that superconductivity vanishes entirely as the phase-winding rate $u \equiv \xi|\nabla\varphi| \to 1$. Assuming that this remains approximately valid in non-bulk materials, and using that $|\nabla\varphi| = |n|/r$ around an $n$-winding vortex, we find that superconductivity vanishes for $r < n\xi$. In other words, find that the core size of a giant vortex scales linearly with its winding number $n$, providing an observational signature of giant vortices that can be seen via STM measurements.

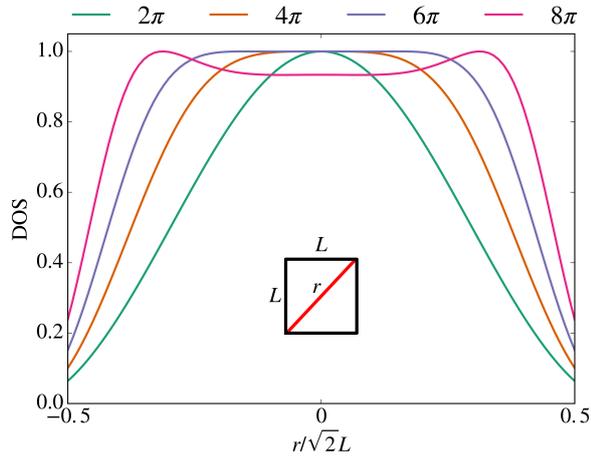

FIG. 4: DOS along the diagonal of the normal metal for various applied current windings $\Phi_I$. Superconductivity is suppressed in the vortex cores, and the normal-state DOS = 1 is recovered.

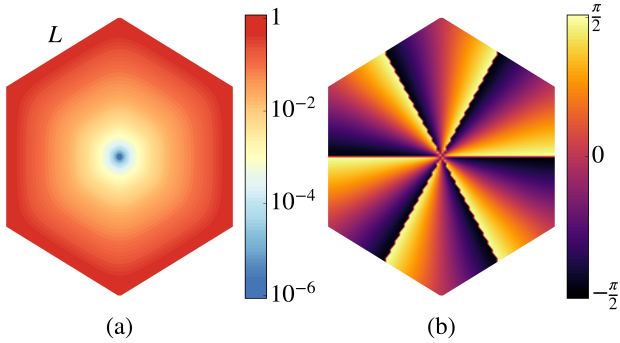

FIG. 5: Hexagonal geometry with side lengths $L = 6\xi$, and an applied current winding $\Phi_I = 6\pi$. (a) The pair correlation $\Psi$, showing a single vortex at the center. (b) The phase $\varphi$ of the pair correlation, demonstrating that it is a giant vortex with winding number $n = +3$.

The vortex patterns of Fig. 3 may be deduced from energy considerations. In general, the kinetic energy of a vortex with a winding number $n$ scales as $n^2$. This is because kinetic energy $E_k \sim v^2$, where $v \sim \nabla\varphi \sim n$ is the velocity of the superconducting condensate. In the supplemental (Sec. I), we solve the linearized Ginzburg-Landau equation near a vortex with winding number $n$, and use this to confirm that the kinetic energy is indeed proportional to $n^2$. Similar $n^2$ dependencies have previously been noted for magnetic vortices in type-II superconductors [22], and these properties are shared by vortices in proximitized non-superconductors [5, 23].

The above provides a simple prescription for predicting the vortex nucleation pattern. When a total vorticity $N$ is introduced to the system, it splits into vortices with individual windings $n_i$ in a way that satisfies $N = \sum_i n_i$. Among all patterns permitted by the symmetries of the geometry, the energetically favored is the one that minimizes $E = \sum_i n_i^2$. Note that $n_i$ can be either positive or negative, allowing for antivortex nucleation.

In the geometry considered so far, off-center vortices can only appear in a square formation without breaking the symmetry of the system, as is seen in Fig. 3. This symmetry constraint explains why it is possible to produce a vortex with winding $n = +2$. A higher winding is, however, not possible because it will always be energetically favorable to introduce four new vortices away from the center, and potentially, an antivortex in the center. Similar results were found for a mesoscopic superconductor in an applied magnetic field [10–12]. The present analysis differs in that the vortex patterns are generated in an intrinsically non-superconducting material solely by an applying an electric current. A regular polygon with a higher symmetry (larger number of sides), will by the same reasoning as above allow for a higher winding at the center, as any alternative will require a larger number of of $n = +1$ vortices to be distributed in a symmetrical fashion. Figure 5 shows the pair correlation function for a hexagonal normal metal surrounded by a superconductor with an applied current equivalent to $\Phi_I = 6\pi$. Here, we find a single vortex of winding $n = +3$. Generally, a regular polygon with $m$ sides allow for a giant vortex with winding up to $n = \lfloor m/2 \rfloor$.

*Conclusion.* We have used microscopic calculations to show that one can induce giant vortices and antivortices in non-superconducting materials in the absence of magnetic fields. We also analyzed the vortex nucleation pattern using arguments of symmetry and energy minimization. Our results open the possibility to study novel topological defects in unusual environments, which do not have to be intrinsically superconducting or exposed to magnetic fields.

*Acknowledgments.* The authors thank Vetle Kjær Risinggård for useful discussions. This work was partly supported by the Research Council of Norway through its Centres of Excellence funding scheme, project number 262633, "QuSpin". J.L. and J.A.O. also acknowledge funding from Research Council of Norway Grant No. 240806.

[1] A. A. Abrikosov, *Journal of Physics and Chemistry of Solids* **2**, 199 (1957).
[2] G. Blatter, M. V. Feigel'man, V. B. Geshkenbein, A. I. Larkin, and V. M. Vinokur, *Rev. Mod. Phys.* **66**, 1125 (1994).
[3] H. Suderow, I. Guillamon, J. G. Rodrigo, and S. Vieira, *Supercond. Sci. Technol.* **27**, 063001 (2014).
[4] D. Roditchev et al., *Nat. Phys.* **11**, 332 (2015).
[5] J.C. Cuevas and F.S. Bergeret, *Phys. Rev. Lett.* **99**, 217002 (2007).
[6] K. Usadel, *Phys. Rev. Lett.* **25**, 507 (1970).
[7] A. Di Bernardo et al., *Phys. Rev. X* **5**, 041021 (2015).
[8] S. Mironov, A. Mel'nikov, and A. Buzdin, *Phys. Rev. Lett.* **109**, 237002 (2012).
[9] H. J. Fink and A. G. Presson, *Phys Rev.* **151**, 219 (1966).
[10] L.F. Chibotaru, A. Ceulemans, V. Bruyndoncx, and V.V. Moshchalkov, *Nature* **408**, 833 (2000).
[11] L.F. Chibotaru, A. Cuelemans, V. Bruyndoncx, and V.V. Moshchalkov, *Phys. Rev. Lett.* **86**, 1323 (2001).
[12] L.F. Chibotaru, A. Ceulemans, G. Teniers, and V.V. Moshchalkov, *Physica C* **369**, 149 (2002).

[13] J. Rammer and H. Smith, *Rev. Mod. Phys.* **58**, 323 (1986).
[14] W. Belzig et al., *Superlattice Microst.* **25**, 1251 (1999).
[15] M.Y. Kupriyanov and V.F. Lukichev, *Sov. Phys. JETP* **67**, 1163 (1988).
[16] N. Schopohl, arxiv:cond-mat/9804064.
[17] M. Amundsen and J. Linder, *Sci. Rep.* **6**, 22765 (2016).
[18] S.V. Bakurskiy et al, *Phys. Rev. B* **88**, 144519 (2013).
[19] J. Romijn, T.M. Klapwijk, M.J. Renne, and J.E. Mooij, *Phys. Rev. B* **26**, 3648 (1982).
[20] I. V. Grigorieva, W. Escoffier, V. R. Misko, B. J. Baelus, F. M. Peeters, L. Y. Vinnikov, and S. V. Dubonos, *Phys. Rev. Lett.* **99**, 147003 (2007).
[21] G. Teniers, L. F. Chibotaru, A. Ceulemans, and V. V. Moshchalkov, *Europhys. Lett.* **63**, 296 (2003).
[22] P.G. de Gennes, *Superconductivity of Metals and Alloys* (W.A. Benjamin, New York, 1966).
[23] F.S. Bergeret and J.C. Cuevas, *J. Low Temp. Phys.* **153**, 304 (2008).

# Supplemental information

Morten Amundsen,[1] Jabir Ali Ouassou,[1] and Jacob Linder[1]

[1]*Center for Quantum Spintronics, Department of Physics, Norwegian University of Science and Technology, NO-7491 Trondheim, Norway*

(Dated: February 24, 2018)

In this supplemental, we derive two sets of equations that are applied in the main manuscript. In Section I, we find an exact solution to the linearized Ginzburg-Landau equation near a vortex with an arbitrary winding number $n$, and show that the kinetic energy is proportional to $n^2$. This is used to explain the vortex configurations in the main manuscript. In Section II, we find a selfconsistent solution to the Usadel equation in a bulk superconductor with a uniform charge current. This is used as a boundary condition in the main manuscript.

## I. ANALYTICAL SOLUTION AROUND A VORTEX CORE

Let us consider a superconducting vortex with a winding number $n$. This means that as we move one counter-clockwise turn around the vortex core, the phase of the superconducting condensate changes by $\Delta\varphi = 2\pi n$. We will here calculate the energy of such a vortex, which in the main manuscript is used to understand what nucleation patterns are energetically favored. To keep the calculations simple and intuitive, we approach the problem using the Ginzburg-Landau formalism. Furthermore, we will assume that the energy of a vortex is dominated by the region close to the vortex core, and that this region exhibits a cylindrical symmetry. Since the energy of a vortex ($n > 0$) and antivortex ($n < 0$) are exactly the same, we focus on $n > 0$.

### A. Linearized Ginzburg-Landau theory

The starting point of the Ginzburg-Landau framework is the free energy density in a superconducting material,[1]

$$E = \alpha|\Psi|^2 + \frac{\beta}{2}|\Psi|^4 + \frac{1}{2m}|(-i\hbar\nabla - 2e\boldsymbol{A})\Psi|^2 + \frac{\boldsymbol{B}^2}{2\mu_0}, \quad (1)$$

where $\Psi$ is the superconducting wavefunction, $\boldsymbol{A}$ is the magnetic potential, $\boldsymbol{B} = \nabla \times \boldsymbol{A}$ is the magnetic field, and we choose $E \equiv 0$ in the normal state. Minimizing the free energy of the system, one arrives at the Ginzburg-Landau equation,[1]

$$\alpha\Psi + \beta|\Psi|^2\Psi + \frac{1}{2m}(-i\hbar\nabla - 2e\boldsymbol{A})^2\Psi = 0. \quad (2)$$

We now introduce some approximations. Firstly, as in the main manuscript, we are interested in current-induced vortices in thin-films, for which there is a negligible magnetic potential $\boldsymbol{A} \approx 0$ in the system. Secondly, we are interested in the behaviour near a vortex core, where the superconducting wavefunction is suppressed $|\Psi| \ll 1$, so that we can linearize the equation. We then obtain an effective Helmholtz equation,

$$\nabla^2\Psi \approx \Psi/\xi_0^2, \quad (3)$$

where $\xi_0 \equiv \sqrt{\hbar^2/2m|\alpha|}$ is the Ginzburg-Landau coherence length. We can parametrize the wavefunction as $\Psi \equiv \psi e^{i\varphi}$, where the amplitude $\psi$ and phase $\varphi$ are real. Substituting this parametrization into the Helmholtz equation, we obtain

$$\nabla^2\psi + 2i(\nabla\psi)(\nabla\varphi) + i\psi(\nabla^2\varphi) - \psi(\nabla\varphi)^2 = \psi/\xi_0^2. \quad (4)$$

This equation can be significantly simplified using the law of charge conservation. The charge current density in a system governed by the Ginzburg-Landau equation is in general:[1]

$$\boldsymbol{J} = \frac{e}{m}[\Psi^*(-i\hbar\nabla - 2e\boldsymbol{A})\Psi + \Psi(+i\hbar\nabla - 2e\boldsymbol{A})\Psi^*]. \quad (5)$$

If we again set $\boldsymbol{A} \approx 0$ and substitute in $\Psi = \psi e^{i\varphi}$,

$$\boldsymbol{J} = \frac{2\hbar e}{m}\psi^2 \nabla\varphi. \quad (6)$$

From this equation for the charge current, combined with the fact that charge current is conserved $\nabla \cdot \boldsymbol{J} = 0$, we conclude:

$$2\psi(\nabla\psi)(\nabla\varphi) + \psi^2\nabla^2\varphi = 0. \quad (7)$$

At any point with a finite wavefunction $\psi \neq 0$, this means that two of the terms on the left-hand side of Eq. (4) have to cancel. This lets us write Eq. (4) as simply:

$$\nabla^2\psi - \psi(\nabla\varphi)^2 = \psi/\xi_0^2. \quad (8)$$

### B. Exact vortex profile

We now focus on the specific case of a vortex with winding $n$, meaning that the total phase-difference around the core is $\Delta\varphi = 2\pi n$. At a distance $r$ from the core, this phase-difference occurs over a length $2\pi r$, yielding an average phase-gradient $|\nabla\varphi| = \Delta\varphi/2\pi r = n/r$. Assuming cylindrical symmetry, we expect the amplitude $\psi$ to only depend on the radius $r$ from the vortex core, so that $\nabla^2\psi \to r^{-1}\partial_r(r\partial_r\psi)$. Together, these observations let us reduce Eq. (8) to an ordinary differential equation for the radial profile $\psi(r)$, which can be written as:

$$r^2\frac{d^2\psi}{dr^2} + r\frac{d\psi}{dr} - \left(n^2 + \frac{r^2}{\xi_0^2}\right)\psi = 0. \quad (9)$$

This is the defining equation for the *modified* or *hyperbolic* Bessel functions $I_n(r/\xi_0)$ and $K_n(r/\xi_0)$. However, whereas the first kind $I_n(r/\xi_0)$ always converges to a finite value as $r \to 0$, the second kind diverges there, and is therefore an unphysical solution. The radial profile of a vortex is therefore:

$$\psi(r) = \psi_0 I_n(r/\xi_0). \quad (10)$$

### C. Asymptotic kinetic energy

In the previous subsection, we found exact solutions of the linearized Ginzburg-Landau equation in the vicinity of a vortex. These are however not straight-forward to use for analytically comparing vortex energies. Physically, we expect the dominant contributions to the kinetic energy to come from the region close to the vortex. This means that we can do a Taylor expansion around the vortex core $r = 0$,

$$I_n(r/\xi_0) = \sum_{m=0}^{\infty} \frac{1}{m!(m+n)!} \left(\frac{r}{2\xi_0}\right)^{2m+n} \quad (11)$$

and focus on the region near the vortex core $r \ll \xi_0$ where the $m = 0$ term becomes the dominant contribution. This gives us the following asymptotic profile for a vortex with winding $n$:

$$\psi(r) \approx \frac{\psi_0}{n!} \left(\frac{r}{2\xi_0}\right)^n. \quad (12)$$

We can now go back to the free energy, and use these solutions to determine the energy associated with each vortex. Let us consider the kinetic energy density $E_k$. In the absence of magnetism, this is just the gradient term in Eq. (1):

$$E_k = \frac{\hbar^2}{2m} |\nabla \Psi|^2. \quad (13)$$

We then switch to polar coordinates $\nabla = \partial_r \boldsymbol{e}_r + r^{-1} \partial_\theta \boldsymbol{e}_\theta$:

$$E_k = \frac{\hbar^2}{2m} \left( |\partial_r \Psi|^2 + r^{-2} |\partial_\theta \Psi|^2 \right). \quad (14)$$

Substituting in the asymptotic solutions $\Psi \sim r^n e^{in\theta}$:

$$E_k = \frac{\hbar^2 |\Psi|^2}{mr^2} n^2. \quad (15)$$

Thus, the kinetic energy of a giant vortex is proportional to $n^2$.

## II. SUPERCONDUCTOR WITH A UNIFORM CURRENT

In the main manuscript, we considered a system consisting of a superconducting wire encircling a normal metal. Although the superconductor was assumed to be thick enough to act as a bulk material, the fact that it also carries a supercurrent means that the propagators are no longer given by the standard BCS solution. In order to use as realistic boundary conditions as possible for that setup, we here solve the Usadel equation analytically for a current-carrying superconductor.

### A. Background theory

In a superconductor, the Usadel equation can be written[2–4]

$$iD\nabla(\hat{g}\nabla\hat{g}) = [\epsilon\hat{\tau}_3 + \hat{\Delta},\, \hat{g}], \quad (16)$$

where $\hat{\tau}_3 = \text{diag}(+1, +1, -1, -1)$, and the gap matrix is defined as $\hat{\Delta} = \text{antidiag}(+\Delta, -\Delta, +\Delta^*, -\Delta^*)$. The superconducting gap can in turn be parametrized as $\Delta = |\Delta|e^{i\varphi}$ where $\varphi \in \mathbb{R}$. The matrices on the left-hand side of the commutator are then:

$$\epsilon\hat{\tau}_3 = \epsilon \begin{pmatrix} +\sigma_0 & 0 \\ 0 & -\sigma_0 \end{pmatrix}, \quad \hat{\Delta} = |\Delta| \begin{pmatrix} 0 & e^{+i\varphi}i\sigma_2 \\ e^{-i\varphi}i\sigma_2 & 0 \end{pmatrix}. \quad (17)$$

The propagator $\hat{g}$ can be written using the $\theta$-parametrization:[4,5]

$$\hat{g} = \begin{pmatrix} +\cosh\theta\,\sigma_0 & e^{+i\chi}\sinh\theta\,i\sigma_2 \\ e^{-i\chi}\sinh\theta\,i\sigma_2 & -\cosh\theta\,\sigma_0 \end{pmatrix}. \quad (18)$$

The parameters $\theta$ and $\chi$ satisfy the particle-hole symmetries $\theta^*(+\epsilon) = -\theta(-\epsilon)$ and $\chi^*(+\epsilon) = \chi(-\epsilon)$.[5] For brevity, we also use the abbreviations $s \equiv \sinh\theta$ and $c \equiv \cosh\theta$. Finally, the self-consistency equation for the gap is:[8]

$$\Delta = N_0 \lambda e^{i\chi} \int_0^{\omega_c} d\epsilon\, \text{Re}[\sinh\theta] \tanh(\epsilon/2T). \quad (19)$$

Comparing this to the parametrization of the gap $\Delta = |\Delta|e^{i\varphi}$, we immediately note that the phases $\varphi = \chi$ must be equal.

### B. Zero current

In the absence of charge currents, we must have a homogeneous solution $\nabla\hat{g} = 0$. Thus, the Usadel equation has to reduce to:

$$[\epsilon\hat{\tau}_3 + \hat{\Delta},\, \hat{g}] = 0. \quad (20)$$

Writing the terms in the commutator explicitly, we get:

$$[\epsilon\hat{\tau}_3,\, \hat{g}] = +2\epsilon \begin{pmatrix} 0 & +se^{+i\varphi}i\sigma_2 \\ -se^{-i\varphi}i\sigma_2 & 0 \end{pmatrix}, \quad (21)$$

$$[\hat{\Delta},\, \hat{g}] = -2|\Delta| \begin{pmatrix} 0 & +ce^{+i\varphi}i\sigma_2 \\ -ce^{-i\varphi}i\sigma_2 & 0 \end{pmatrix}. \quad (22)$$

From this, we can extract the scalar equation $\epsilon s - |\Delta|c = 0$, which yields the standard BCS solution $\theta = \text{atanh}(|\Delta|/\epsilon)$.

### C. Uniform current

Before we attempt to solve the Usadel equation in a current-carrying superconductor with $\partial_z \hat{g} \neq 0$, let us try to constrain the allowed position-dependence of our parameters $\theta$ and $\varphi$. One such condition can be found from the density of states,

$$N = \frac{1}{2} N_0 \,\text{Re}\, \text{Tr}[g] = N_0\, \text{Re}[\cosh\theta]. \quad (23)$$

For a bulk superconductor carrying a uniform current, we insist that the density of states is uniform as well, i.e. that $\partial_z N = 0$. Using the chain rule, we can rewrite this condition as follows:

$$(\partial_z \theta)(\partial_\theta \text{Re}[\cosh\theta]) = 0. \quad (24)$$

Thus, we may either have $\partial_z \theta = 0$ or $\partial_\theta \text{Re}[\cosh\theta] = 0$. Since $\theta$ is a direct function of energy, the latter is equivalent to the density of states being energy-independent, which we know is false for a superconductor. Thus, we conclude that $\partial_z \theta = 0$.

Now that we know $\partial_z \theta = 0$, differentiating $\hat{g}$ is quite easy:

$$\partial_z \hat{g} = i\partial_z \varphi \begin{pmatrix} 0 & +e^{+i\varphi} si\sigma_2 \\ -e^{-i\varphi} si\sigma_2 & 0 \end{pmatrix}. \quad (25)$$

Multiplying by $\hat{g}$ from the left, we then obtain:

$$\hat{g}\partial_z \hat{g} = i\partial_z \varphi \begin{pmatrix} s^2 \sigma_0 & e^{+i\varphi} cs\, i\sigma_2 \\ e^{-i\varphi} cs\, i\sigma_2 & -s^2 \sigma_0 \end{pmatrix}. \quad (26)$$

Another constraint can then be found from the spectral current,

$$j_z = \frac{1}{4} j_0 \, \text{Tr}[\hat{\tau}_3 \hat{g} \partial_z \hat{g}]. \quad (27)$$

Substituting in the expression for $\hat{g}\partial_z\hat{g}$ above, we find that $j_z/j_0 = is^2 \partial_z \varphi$. But insisting that the divergence $\partial_z j_z = 0$, and keeping in mind that $\partial_z s = 0$ because we determined that $\partial_z \theta = 0$ above, this gives us the constraint $\partial_z^2 \varphi = 0$. One might however argue that perhaps the spectral current does not have to be conserved, since charge conservation only requires that the *integral* of the spectral current above is position-independent. However, for a *uniform current-carrying superconductor*, we can safely insist that the spectral current be constant as well.

Now that we have the additional constraint $\partial_z^2 \varphi = 0$, it is straight-forward to differentiate $\hat{g}\partial_z\hat{g}$:

$$\partial_z(\hat{g}\partial_z\hat{g}) = (i\partial_z\varphi)^2 \begin{pmatrix} 0 & +e^{+i\varphi} cs\, i\sigma_2 \\ -e^{-i\varphi} cs\, i\sigma_2 & 0 \end{pmatrix}. \quad (28)$$

This defines the left-hand side of the Usadel equation. Combining the above with the rest of the Usadel equation, we find the following equation from the off-diagonal parts:

$$iD(i\partial_z\varphi)^2 cs = 2\epsilon s - 2|\Delta|c. \quad (29)$$

We will now normalize everything with respect to the zero-current gap $\Delta_0$, so that $|\Delta| \equiv \delta \Delta_0$ and $\epsilon \equiv E\Delta_0$. Furthermore, we define a phase-winding rate $u^2 \equiv D(\partial_z\varphi)^2/\Delta_0$. Thus:

$$Es - \delta c + i(u^2/2)cs = 0. \quad (30)$$

Note that since the diffusion constant can be written $D = \Delta_0 \xi^2$, we could also write $u = \xi \partial_z \varphi$, which means that this parameter basically measures the phase-winding per coherence length. By substituting the hyperbolic identity $c = \sqrt{1+s^2}$ into Eq. (30), the resulting 4th-order algebraic equation in $s$ can easily be solved to provide the analytical solution. However, for practical reasons we here pursue a numerical approach.

### D. Non-selfconsistent solution

In order to solve Eq. (30), it is convenient to reparametrize the equation using the following mapping, where $\Theta(E)$ is an unknown function of energy:[6,7]

$$c = \frac{E}{\sqrt{E^2 - \Theta^2}}, \qquad s = \frac{\Theta}{\sqrt{E^2 - \Theta^2}}. \quad (31)$$

Note that this parametrization manifestly satisfies the identity $c^2 - s^2 = 1$. Substituting the above into Eq. (30) and rearranging, we find that the Usadel equation can be rewritten as:

$$\Theta = \frac{\delta}{1 + u^2/2\sqrt{\Theta^2 - E^2}}. \quad (32)$$

In the absence of currents $u = 0$, we get a trivial solution $\Theta = \delta$. For a finite phase-winding rate $u$, it takes the form of a fixpoint iteration equation, and can be solved using Newton's method.

In addition to the above equation for $\Theta$, we need to determine the superconducting phase $\varphi$. However, in the previous subsection, we discovered that $\partial_z^2 \varphi = 0$. This means that the phase $\varphi$ has to be a linear function of position. Furthermore, since the reference-point for the superconducting phase is arbitrary, we can define $\varphi(0) \equiv 0$. Thus, the phase $\varphi$ can be expressed as:

$$\varphi(z) = uz/\xi. \quad (33)$$

For small currents, one can safely assume that the gap is nearly the same as for zero current, meaning that $\delta \approx 1$. However, in general, this fixpoint equation has to be accompanied by a selfconsistency equation for the current-dependent gap factor $\delta$.

### E. Selfconsistent solution

Let us now revisit the selfconsistency equation for the gap, using the $\Theta$-parametrization from the previous subsection. We normalize the energy $E \equiv \epsilon/\Delta_0$, gap $\delta \equiv |\Delta|/\Delta_0$, Debye cutoff $\Omega_c \equiv \omega_c/\Delta_0$, and temperature $\tau \equiv T/T_c$. Furthermore, the cutoff is in general related to the BCS coupling strength by $\Omega_c = \cosh(1/N_0\lambda)$, while the gap and critical temperature are related by the BCS ratio $\Delta_0/T_c = \pi/e^\gamma$, where $\gamma$ is the Euler-Mascheroni constant.[8] Combining all of these remarks, Eq. (19) for the current-dependent gap may be written as:

$$\delta = \frac{1}{\text{acosh}\,\Omega_c} \int_0^{\Omega_c} \! dE \, \text{Re}\left(\frac{\Theta}{\sqrt{E^2 - \Theta^2}}\right) \tanh\left(\frac{\pi}{2e^\gamma} \frac{E}{\tau}\right). \quad (34)$$

In general, the selfconsistent problem is solved in two steps. First, we guess that the solution is $\Theta(E) = 1$ and $\delta = 1$. For each energy in a discretized range from $E = \Omega_c$ to $E = 0$, one solves Eq. (32) for $\Theta(E)$ using Newton's method. The solutions are substituted into Eq. (34), which is integrated to find a new estimate for $\delta$. This procedure is repeated until convergence.

## III. GIANT VORTICES IN ASYMMETRIC GEOMETRIES

Giant vortices are inherently unstable and will seek to split into single vortices unless hindered from doing so. For the systems under consideration, the giant vortices are maintained due to symmetry constraints. It is therefore interesting to investigate how, for instance, Fig. 3(b) in the manuscript reacts to a small deviation from perfect symmetry. We do so by introducing a small perturbation $\varepsilon$ of the aspect ratio $\alpha$ by defining $\alpha = 1 + \varepsilon$, thereby making the system rectangular. The results are shown in Fig. 1, from which it is seen that vortices do indeed split as $\varepsilon$ is increased, but this splitting occurs in a continuous way, and the resulting vortex pair remains within close proximity to the location of the original giant vortex for a deviation of up to $\varepsilon = 1\%$. This means that the giant vortex can be stabilized against small deviations in the geometry by placing a pinning potential at this position,[9] or by forcing the split vortices together by fine tuning the applied currents. To reduce the influence of unintended asymmetry, it is recommended to use a superconductor with as large a coherence length as possible. Choosing for instance aluminium, one gets an estimated diffusive coherence length of $\xi \simeq 100$ nm. For a square geometry with side lengths $L = 12\xi$, a deviation of $\varepsilon = 1\%$ then corresponds to $\Delta L \simeq 12$ nm, which is an experimentally achievable level of accuracy.[10]

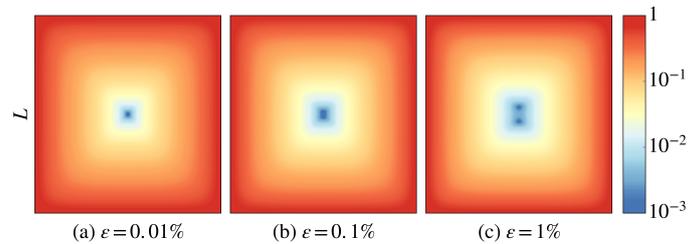

(a) $\varepsilon = 0.01\%$ (b) $\varepsilon = 0.1\%$ (c) $\varepsilon = 1\%$

FIG. 1: Vortex patterns for an applied current winding of $\Phi_I = 4\pi$, with increasing aspect ratio deviation $\varepsilon$.

[1] K. Fossheim and A. Sudbø, *Superconductivity: Physics and Applications* (John Wiley & Sons, Chichester, 2004), Sec. 4.7.
[2] K. Usadel. *Phys. Rev. Lett.* **25**, 507 (1970).
[3] J. Rammer and H. Smith. *Rev. Mod. Phys.* **58**, 323 (1986).
[4] W. Belzig, F. K. Wilhelm, C. Bruder, G. Schön, and A. D. Zaikin. *Superlattices and Microstructures* **25**, 1251 (1999).
[5] J.P. Morten, Master thesis, NTNU, 2003.
[6] S.V. Bakurskiy, N.V. Klenov, I.I. Soloviev, M.Yu. Kupriyanov, and A.A. Golubov. *Phys. Rev. B* **88**, 144519 (2013).
[7] J. Romijn, T.M. Klapwijk, M.J. Renne, and J.E. Mooij. *Phys. Rev. B* **26**, 3648 (1982).
[8] S.H. Jacobsen, J.A. Ouassou, and J. Linder. *Phys. Rev. B* **92**, 024510 (2015).
[9] I. V. Grigorieva, W. Escoffier, V. R. Misko, B. J. Baelus, F. M. Peeters, L. Y. Vinnikov, and S. V. Dubonos. *Phys. Rev. Lett.* **99**, 147003 (2007).
[10] W. Hu, K. Sarveswaran, M. Lieberman, and G. H. Bernstein. *J. Vac. Sci. Technol. B* **22**, 1711 (2004).